\documentclass{aa}  
\usepackage{graphicx}
\usepackage{natbib}

\usepackage{txfonts}
\begin{document}
 \title{Quiescent and flaring X-ray emission from the nearby M/T dwarf binary SCR~1845-6357}

   \author{J. Robrade
          \and
          K. Poppenhaeger
\and
          J.H.M.M. Schmitt
          }


        \institute{Universit\"at Hamburg, Hamburger Sternwarte, Gojenbergsweg 112, D-21029 Hamburg, Germany\\
       \email{jrobrade@hs.uni-hamburg.de}
             }

   \date{Received 5 November 2009 / Accepted...}

 
  \abstract
{}
   {X-ray emission is an important diagnostics to study magnetic activity in very low mass stars that are presumably fully convective and have an effectively neutral photosphere.}
   {We investigate an XMM-Newton observation of SCR~1845-6357, a nearby, ultracool M\,8.5\,/\,T\,5.5 dwarf binary. The binary is unresolved in the XMM detectors, however the X-ray emission is very likely from the M~8.5 dwarf. We compare its flaring emission to those of similar very low mass stars and additionally present an XMM observation of the M\,8 dwarf VB~10.}
   {We detect quasi-quiescent X-ray emission from SCR~1845-6357 at soft X-ray energies in the 0.2\,--\,2.0\,keV band,
as well as a strong flare with a count rate increase of a factor of 30 and a duration of only 10~minutes.
The quasi-quiescent X-ray luminosity of $\log L_{\rm X} = 26.2$~erg/s and the
corresponding activity level of $\log L_{\rm X}$/$L_{\rm bol}= -3.8$ point to a fairly active star.
Coronal temperatures of up to 5~MK and frequent minor variability support this picture.
During the flare, that is accompanied by a significant brightening in the near-UV, plasma temperatures 
of 25\,--\,30~MK are observed and an X-ray luminosity of $L_{\rm X} = 8 \times 10^{27}$~erg/s is reached.}
   {SCR~1845-6357 is a nearby, very low mass star that emits X-rays at detectable levels in quasi-quiescence, implying the existence of a corona. The high activity level, coronal temperatures and the observed large flare point to a rather active star, despite its estimated age of a few Gyr.}

   \maketitle
%

\section{Introduction}

The source \object{SCR J1845-6357} (DENIS-P J184504.9-635747, 2MASS J18450541-6357475) - hereafter SCR~1845 -
was discovered as a nearby, very red, high-proper motion object in the Super\,COSMOS RECONS (SCR) survey \citep{hambly04,hen04};
later trigonometric parallax measurements confirmed its proximity ($3.5 \pm 0.3$~pc) to the Sun \citep{dea05}.
Further measurements improved this result, leading to a distance of $3.85 \pm 0.02$~pc \citep{hen06} and making
SCR~1845 one of the closest stellar systems containing a brown dwarf,
comparable in distance to the famous T~dwarf binary orbiting $\epsilon$~Indi \citep{mcc04}.
Follow-up observations performed in 2005 had shown SCR~1845 to be a late-type dwarf binary that consists 
of an M\,8.5 dwarf with a T\,5.5 dwarf secondary, as deduced from
IR observation with the VLT/NACO \citep{bil06}. This finding makes SCR~1845 the first late M/T dwarf binary system discovered.
The substellar companion is an IR-bright, methane rich brown dwarf and
with a separation of 4.5~AU (1.1\arcsec) in a rather close orbit around its host star.
The age of the system (0.1\,--\,10~Gyr) and the mass of the companion (9\,--\,65~M$_{\rm Jup}$) were, however, only poorly constrained by these data.

In a further work devoted to SCR~1845, \cite{kasp07} present additional IR imaging data
and low-resolution spectroscopy, again obtained with VLT/NACO.
These authors confirm the previous findings and
constrain the mass of the secondary to 40\,--\,50~M$_{\rm Jup}$, thus ruling out a giant planet.
They further estimate the age of the system to 1.8\,--\,3.1~Gyr, consequently SCR~1845 is not particularly young. 
The spectrum of the primary component is well described by an M\,8.5 dwarf with
$T_{\rm eff}\approx 2600$~K, $M\approx 0.09$~M$_{\odot}$ and approximately solar metallicity. 
Adopting their J band magnitude (J~$=9.58 \pm 0.02$)
and bolometric corrections from
\cite{reid01}, we derive a bolometric luminosity of $L_{\rm bol} = 1.0 \pm 0.03 \times 10^{30}$~erg/s for SCR~1845.

Magnetic activity phenomena in the outer atmospheric layers of late-type, very low mass stars are remarkable, 
since these stars are generally assumed to be fully convective so that a solar-type dynamo is not expected to operate.
Furthermore, their cool atmospheres ($T_{\rm eff} \lesssim 2500$\,K) should be virtually neutral, 
leading to a high electric resistivity that inhibits the transport of magnetic energy through the photosphere
and consequently also magnetic activity.
Among other diagnostics like H$\alpha$, X-ray emission can put strong constraints on the possible activity generating mechanisms
and indeed, recent X-ray observation indicate that many ultracool dwarf stars are actually quite active as 
expressed by their $L_{\rm X}$/$L_{\rm bol}$ ratio \citep[see e.g.][]{rob09}.
Also less active very late-type stars exist, the M\,8 dwarfs \object{VB 10} with $\log L_{\rm X}$/$L_{\rm bol}\approx -5$ is a prominent example
\citep{fle03, ber08b}.
Activity finally drops in the L~dwarf regime where older brown dwarfs with $T_{\rm eff} \lesssim 2300$~K reside \citep[see e.g.][]{moh03, aud07}.
Thus at the very end of the stellar main sequence the full range of X-ray activity levels as observed for more massive magnetically active stars, 
i.e. $\log L_{\rm X}$/$L_{\rm bol}= -3~...~-7$, is likely also present; however,
high activity levels persist in late-type dwarfs much longer than in solar-like stars.
The presence of significant magnetic activity in these objects requires the presence of an efficient operating dynamo mechanisms,
for example an $\alpha^{2}$ or turbulent dynamo.

No reports of magnetic activity on SCR~1845 have so far been presented in the literature.
It was neither known to emit X-ray or radio emission nor
does it appear in H$\alpha$ or magnetic field surveys of ultracool dwarfs.

In this paper we present an X-ray observation of the ultracool dwarf binary SCR~1845,
where we detect flaring as well as quasi-quiescent X-ray emission. The structuring of our paper is as follows;
in Sect.\,\ref{ana} we describe the observation and data analysis, in Sect.\,\ref{res} we present our results, discuss flares on very low mass stars
in Sect.\,\ref{flare} and summarize our findings in Sect.\,\ref{sum}.

\section{Observations and data analysis}
\label{ana}

SCR~1845 was observed by {\it XMM-Newton} in September 2008 for approximately 20\,ks (Obs.ID 0551022901);
data analysis was carried out with the standard XMM software, the Science Analysis System (SAS) version~9.0 \citep{sas}.
{\it XMM-Newton} carries three X-ray CCD cameras with moderate spectral resolution, as well as two X-ray grating spectrometers providing
higher spectral resolution. However,
the detected signal in the RGS (Reflection Grating Spectrometer)
detectors is rather weak; while we clearly see a few photons
from prominent emission lines, e.g. \ion{O}{viii} at 19~\AA\, and \ion{O}{vii} at 21.6\,--\,22.1~\AA, the RGS spectra are
not suitable for a quantitative analysis. Thus we consider only X-ray 
data taken with the EPIC (European Photon Imaging Camera), consisting of two MOS and one PN detector; all detectors
operated in the 'Full Frame' mode with the medium filter. The PN is the most sensitive instrument and primarily used for our analysis.
The EPIC detectors are temporally affected by high background levels, 
thus to minimize contamination we extract photons from a 25\,\arcsec~circular region around
the source position and restrict the analysis to the 0.2\,--\,2.0 keV band, where most of the source photons were detected.
Only during a large flare occurring in our observation, photons of higher energies are present and
a harder band (0.2\,--\,3.0~keV) is used for its study.
The background was taken from close-by regions on the same CCD that contains SCR~1845.
We additionally verified our findings by applying standard selection criteria, which exclude the respective 
high-background time intervals.
Since we are dealing with a high proper motion object, we further
derive the exact X-ray position with the source detection algorithm 'edetect\_chain'.

Spectral analysis was performed with XSPEC~V12.3 \citep{xspec}, and we
use multi-temperature plasma models (APEC) with elemental abundances relative to solar values as given by \cite{grsa}. 
We note that the applied metallicity is interdependent with the emission measure, and different
combinations of both parameters lead to very similar results. 
Due to the proximity of the target, interstellar absorption is negligible and
not required in the modelling of our X-ray data.

\begin{figure}[t]
\begin{center}
\includegraphics[width=29mm]{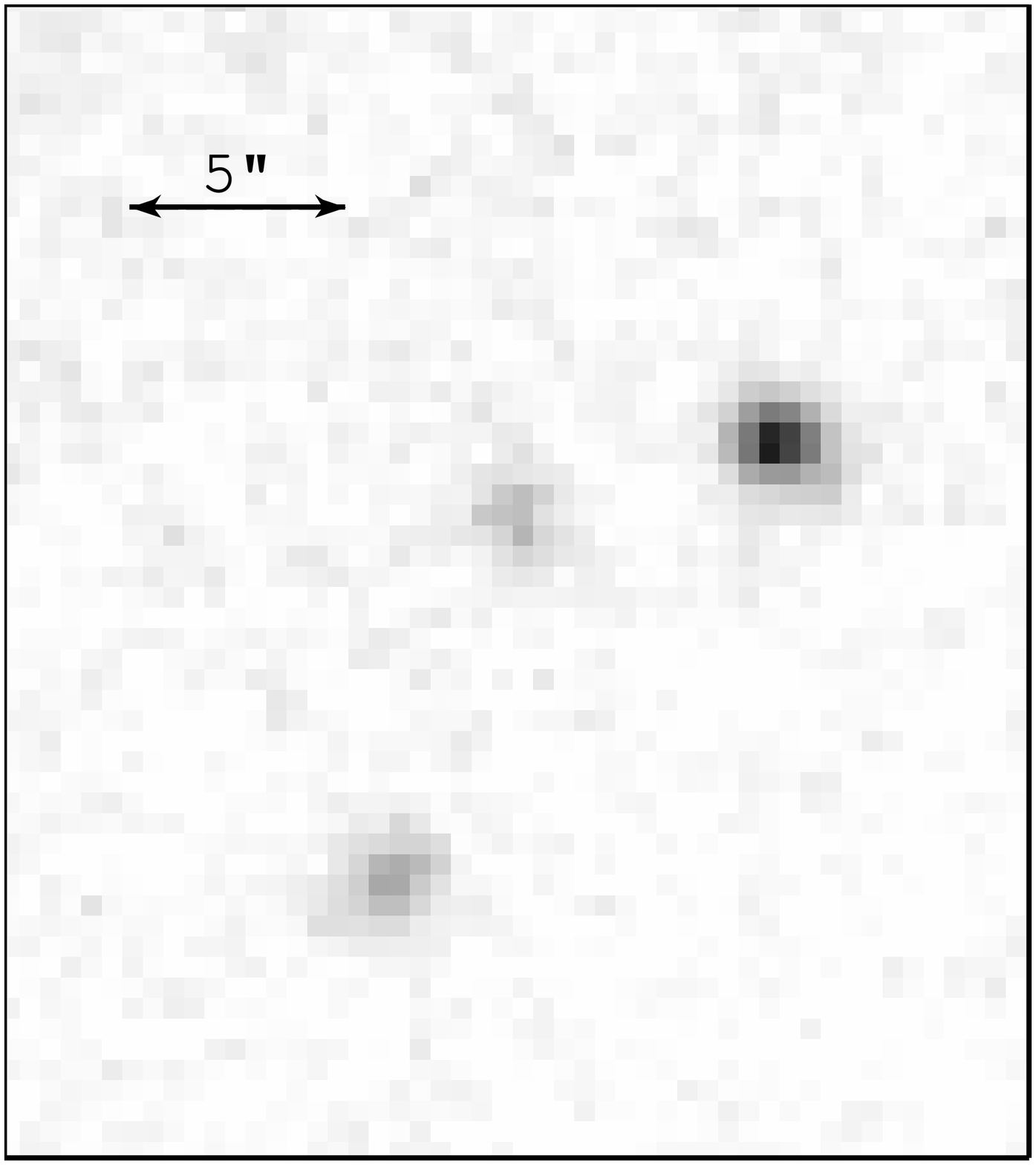}
\includegraphics[width=29mm]{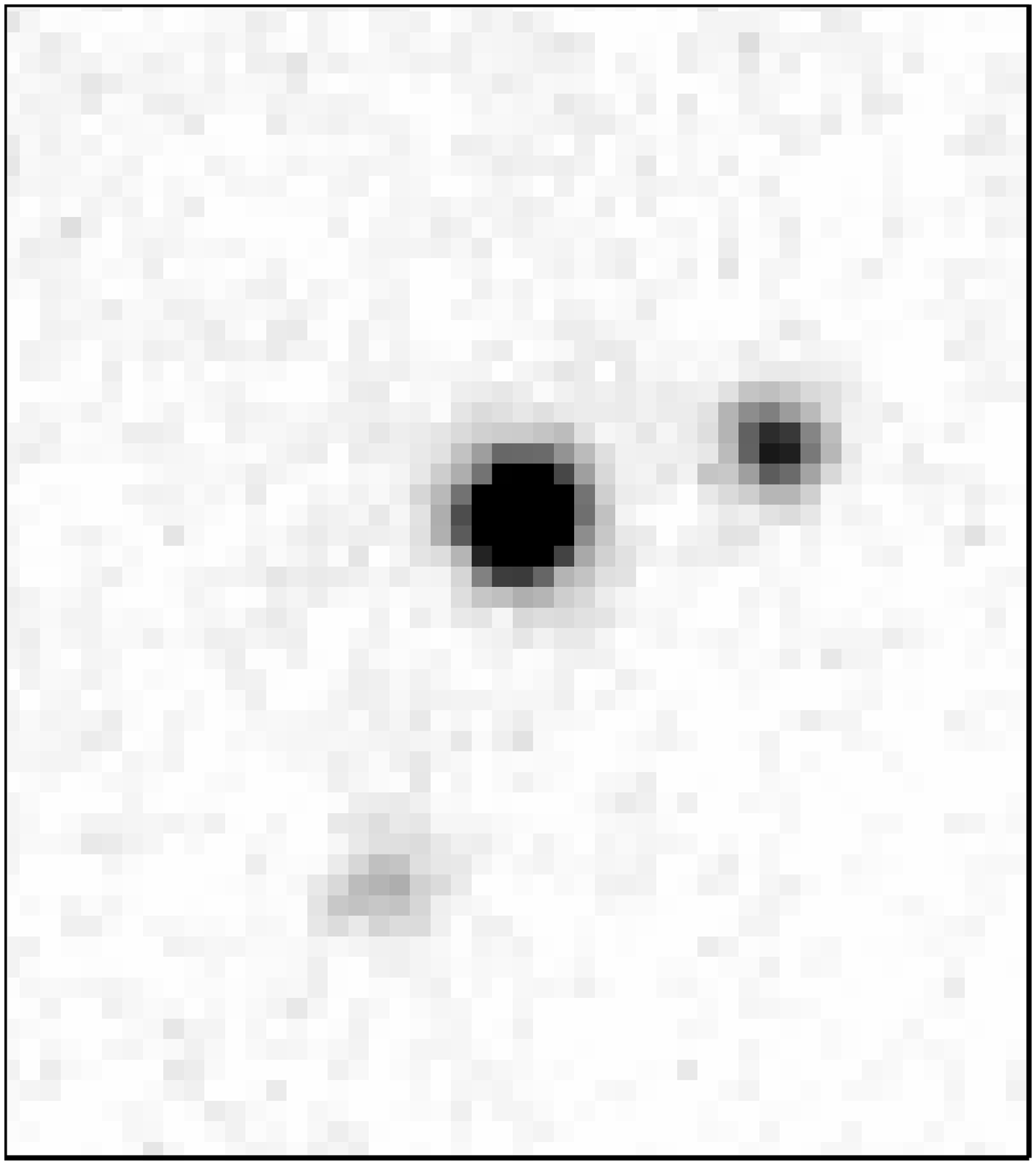}
\includegraphics[width=29mm]{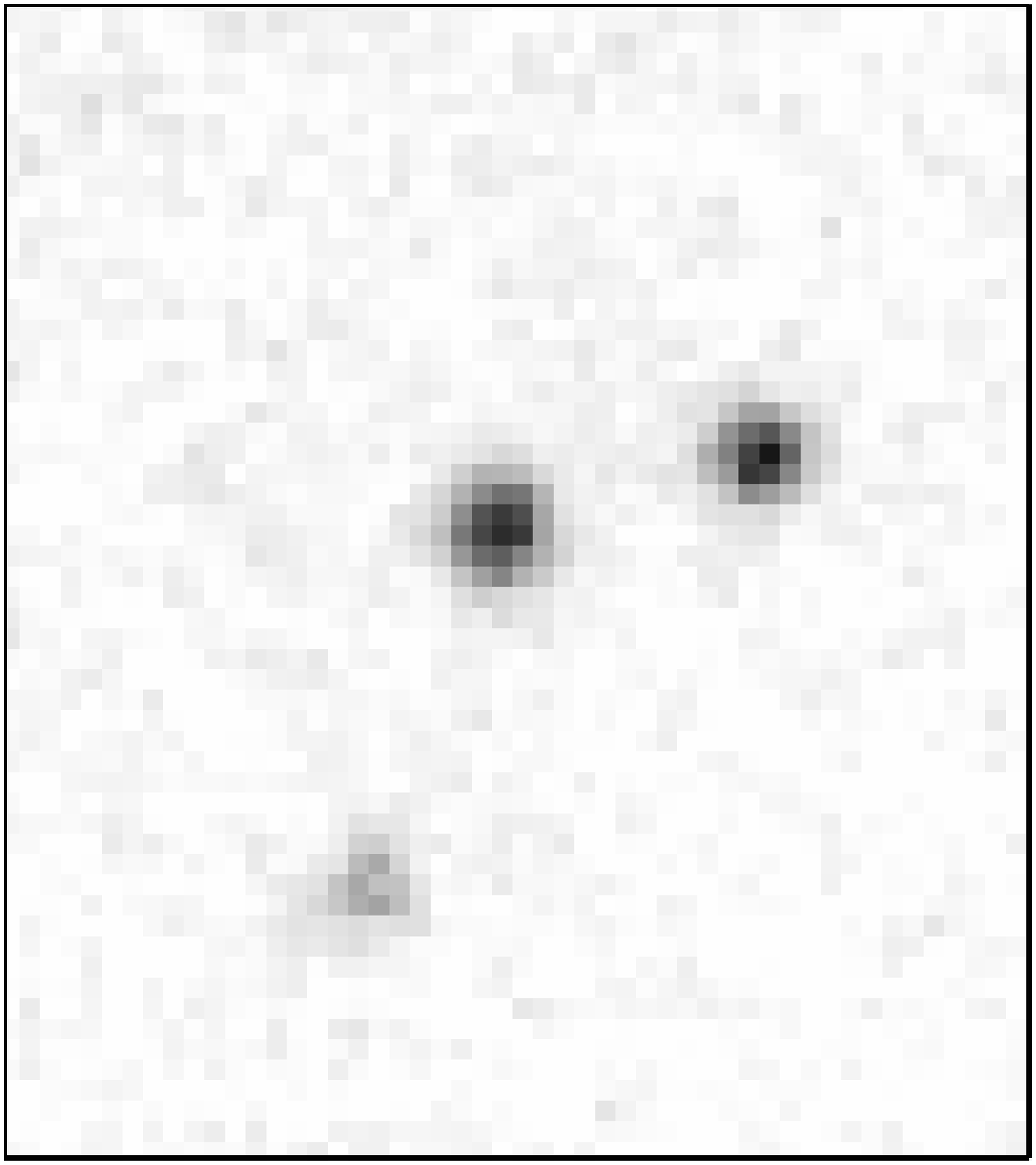}
\caption{\label{opt}OM U band images showing the pre-flare, flare and post-flare state of SCR~1845 (central object, see text for details).}
\end{center}
\end{figure}

The OM (Optical Monitor), an onboard optical/UV telescope, was operated in the imaging mode in the U-band (300\,--\,400~nm, $\lambda_{\rm eff}=344$~nm) 
and took five images with exposure times of 3.3~ks each, 
hence no detailed optical light curve is available. 
Nevertheless, the X-ray flare is clearly accompanied by a significant UV/optical brightening
as can be seen in Fig.~\ref{opt}, where 
we show the pre-flare, flare and post-flare phase of SCR~1845 (center).
In the exposure that covers the flare, SCR~1845 is about 2.6~mag brighter than in the pre-flare
exposure. The other two objects visible in the images,
DENIS J184507.7-635740 (upper right) and USNO-B 0260-0695925 (lower left),
are not in common proper motion with SCR~1845.

\begin{table} [ht!]
\setlength\tabcolsep{4pt}
\begin{center}
\caption{\label{log}X-ray detection of SCR~1845, PN data.}
\begin{tabular}{lll}\hline\hline
Obs.Time & X-ray Position & IR Pos.\\\hline
06 Sept. 08 & RA~~ 18 45 08.63 ($\pm$ 0.25 \arcsec)  & 18 45 08.65 \\
18.5 ks & DEC -63 57 41.6 ($\pm$ 0.25 \arcsec) & -63 57 41.4\\\hline
\end{tabular}
\end{center}
\end{table}

\section{Results}
\label{res}

An X-ray source is clearly detected at the expected position of SCR~1845. The X-ray position coincides very
well with the expected position for epoch 2008.7 coordinates 
as calculated from the known proper motion and the 2MASS position.
Both the positional error and the offset are about 0.3\arcsec~and no other X-ray source is located at distances 
of 1\arcmin~around the position of SCR~1845; thus
the identification is unambiguous.
Additionally the soft X-ray spectra and the observed flare make an unknown extra-galactic source very unlikely.
The source detection parameters are summarized in Table\,\ref{log}.

The system is not resolved and appears like a point-source in the XMM data. While a small contribution from the brown dwarf
to the overall X-ray emission cannot be excluded, it is very likely dominated by the late-type star.
We thus assume in the following
that the detected X-rays originate exclusively from the M\,8.5 dwarf.

\subsection{Light curve analysis}

\begin{figure}[t]
\includegraphics[width=92mm]{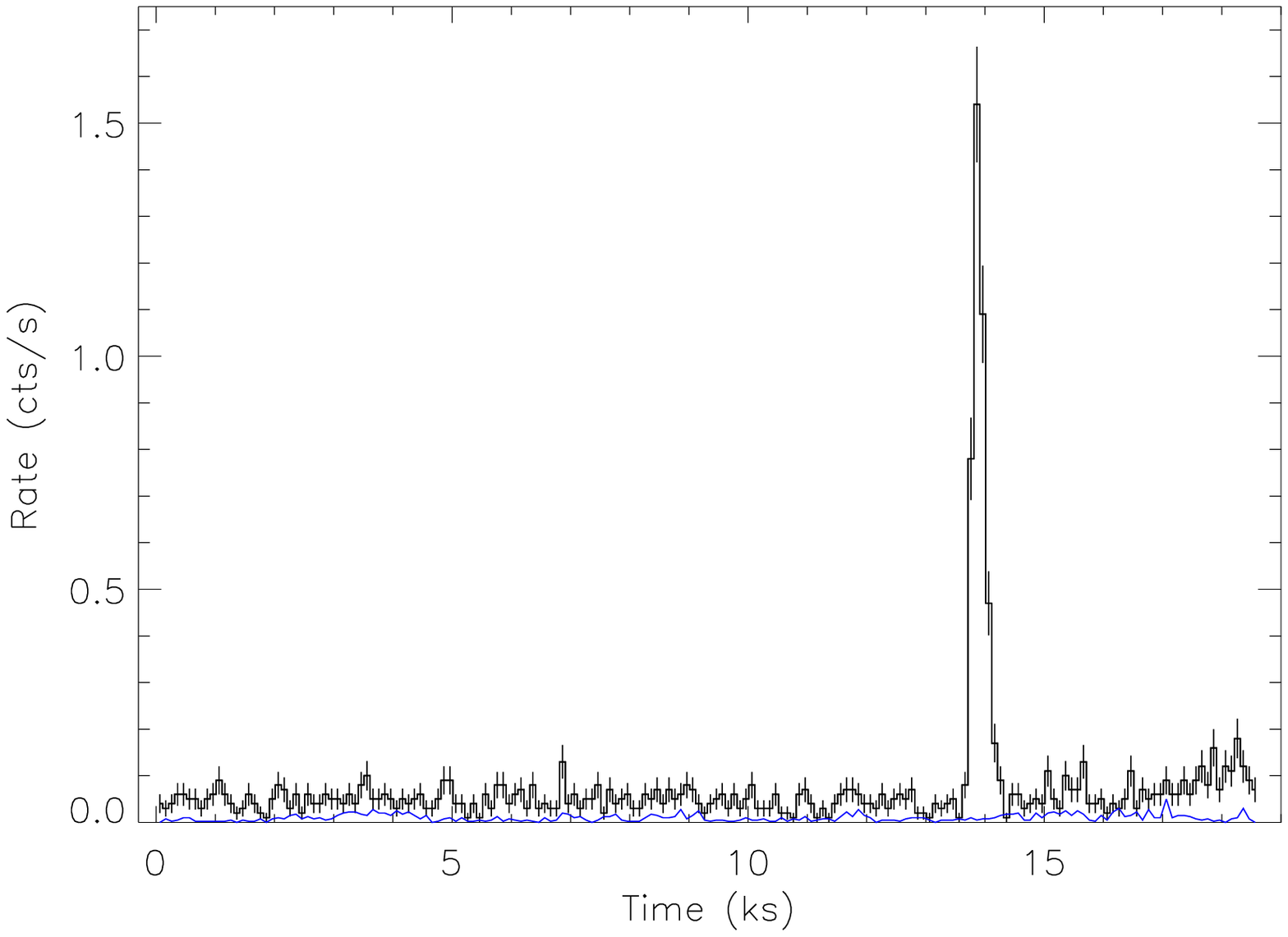}

\vspace*{-6.2cm}
\hspace*{1.3cm}
\includegraphics[width=33mm]{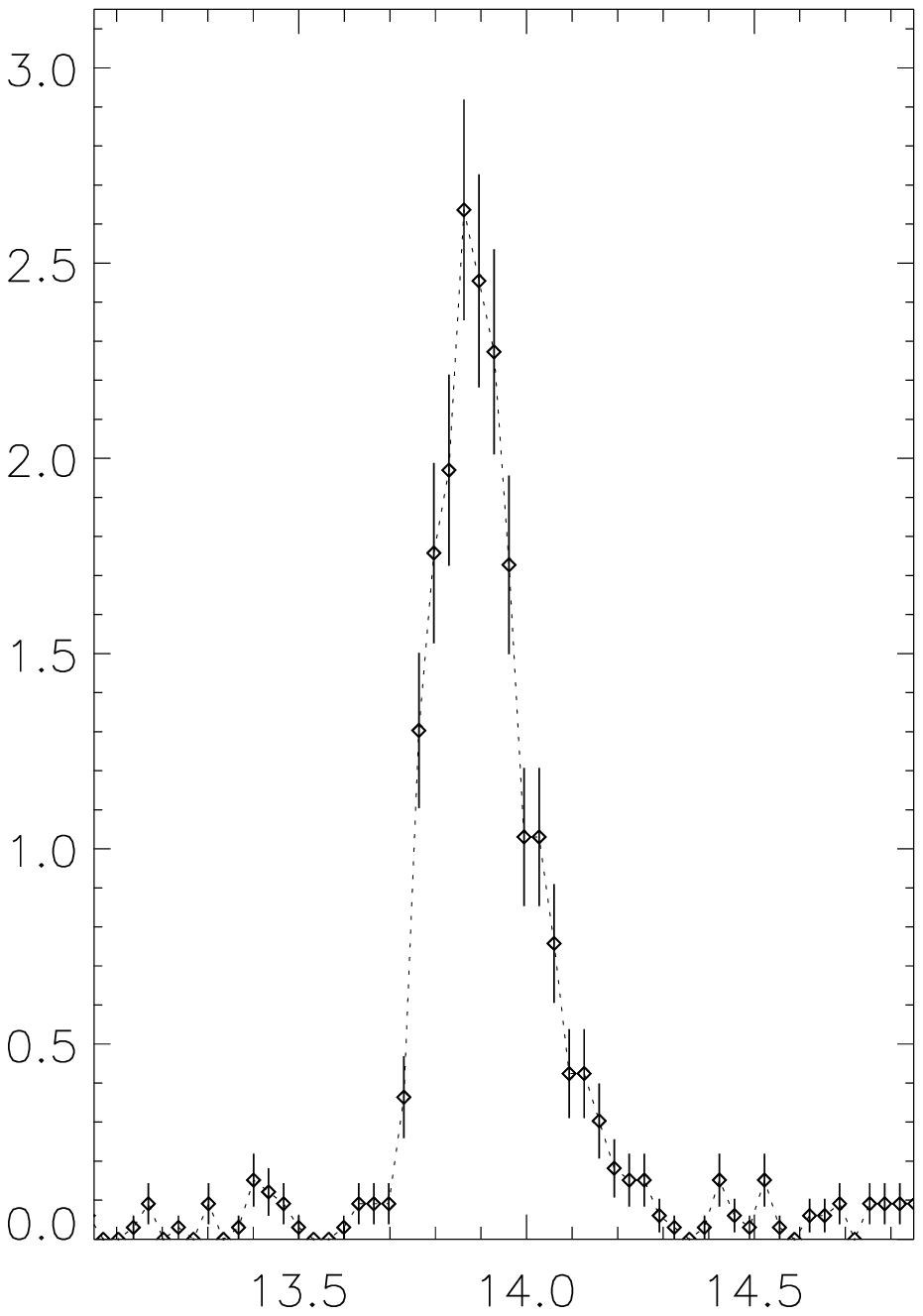}
\vspace*{1.2cm}

\includegraphics[width=92mm]{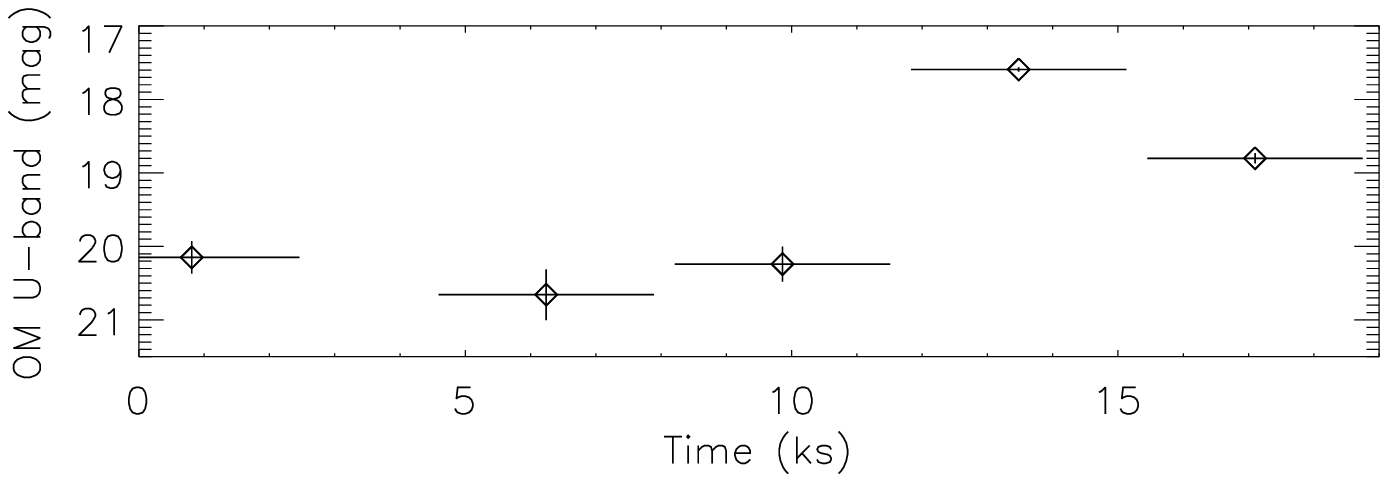}
\vspace*{0.1cm}
\caption{\label{lc}{\it Top:} Light curve of SCR 1845 obtained from PN data with 100~s binning
and corresponding background level (lower blue curve).
{\it Inset}: Zoom of the 0.5~h covering the large flare (EPIC data, 33~s binning).
{\it Bottom:} U-band brightness of the OM exposures.}
\end{figure}

In Fig.\,\ref{lc} we show the X-ray light curve of SCR~1845, which
reveals a strong flare in the second half
of the observation and quasi-quiescent emission over the total exposure time.
Light curves were obtained from the PN data in the 0.2\,--\,2.0~keV band and binned to 100~s
and are plotted separately for the source region and for the background.
The black histogram in the upper panel is the count rate from the source with
1\,$\sigma$ errors, the blue curve is the corresponding background level.
As can be seen, the remaining background is negligible. 
For comparison we further show in the bottom panel of Fig.\,\ref{lc} the U-band brightness averaged over the respective OM exposure.

The X-ray brightness of SCR~1845 is never constant, thus
we call this level quasi-quiescent. During the quasi-quiescent phase the average count rate is 0.05~cts/s.
Variability of up to a factor of about two in count rate on timescales of several minutes is observed throughout the whole exposure.
This behavior points to persistent minor activity on SCR~1845.
An even stronger deviation of the quasi-quiescent level is seen towards
the end of the observation, where 
the average count-rate increases to a value of about 0.1~cts/s.
This additional X-ray emission is no afterglow of the flare; the count-rate drops to the pre-flare level rather
quickly after the event. The flux increase is presumably related to the appearance
of new active regions e.g. due to rotation, or enhanced activity in existing active regions.

The most prominent feature of the X-ray light curve is the large flare;
to investigate the short term behavior during the flare in greater detail we use the combined EPIC, i.e. PN+MOS data.
The inset of Fig.\,\ref{lc} shows the X-ray light curve of the 30~min time interval covering the flare
with a three times higher time resolution (33~s binning).
The peak count rate of the flare is roughly 30\,--\,40 times higher than the average quasi-quiescent level
observed before the flare. 
The strong, burst-like flare has a rise time of about 150~s and an exponential decay time of about 130~s;
thus its light curve has a rather symmetric, triangular shape.
The complete flare event appears and finishes in only ten minutes 
and no elevated flux level is observed before or after the flare.
This behavior already points to a quite compact structure as origin of the flare event.

\subsection{Spectral modeling}

\begin{figure}[t]
\includegraphics[width=50mm,angle=-90]{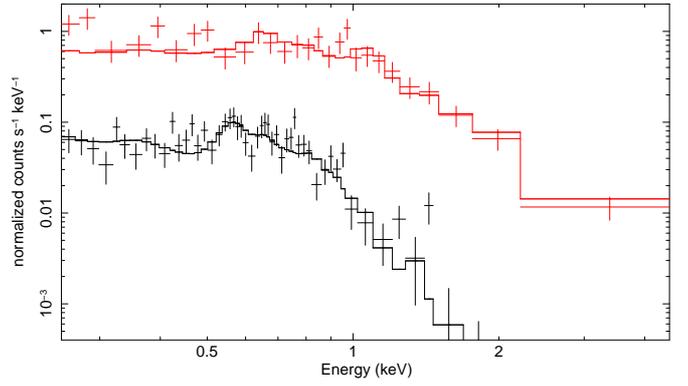}
\caption{\label{spec}PN spectra of the quasi-quiescent phase (black) and the flare (red) with applied spectral models.}
\end{figure}

To determine the spectral properties of the X-ray emission from SCR~1845 we study the quasi-quiescent corona and the overall flare emission
with the PN detector; 
the flare phase covers the 13.7\,--\,14.3~ks time interval.
We applied several spectral models and find that a two-temperature model with solar abundances 
describes the data satisfactorily.
For the flare a three temperature model results in a slightly better fit, but also in rather unconstrained components.
Taking coronal abundances as a free parameter does not significantly improve the fit, thus we adopted solar
abundances for modeling; note however,
that the absolute value of the metallicity can only be poorly constrained with the data.
The results from the spectral modeling are summarized in Table~\ref{fit}. In Fig.~\ref{spec} we show
the PN spectra extracted during quasi-quiescence and flaring with the respective applied model.

\begin{table} [ht!]
\setlength\tabcolsep{5pt}
\begin{center}
\caption{\label{fit} Spectral modeling results derived from the PN data.}
\begin{tabular}{lrrrl}\hline\hline
Par.  & QQ& Flare & Flare Peak & Unit\\\hline
T1& 0.13$\pm$0.02 & 0.35$\pm$0.05 & 0.61$\pm$0.25 & keV\\
EM1&5.4$\pm1.6$ & 36$\pm12$ & 48$\pm22$ &$10^{48}$cm$^{-3}$ \\
T2&0.34$\pm$0.03 & 1.96$\pm$0.42 & 2.46$\pm$0.72  & keV\\
EM2& 4.2$\pm0.9$& 159$\pm35$ & 415$\pm98$  & $10^{48}$cm$^{-3}$\\\hline
$\chi^2_{red}${\tiny(d.o.f.)}& 1.3 (51) & 1.1 (20)& 1.1 (11)&\\\hline
L$_{\rm X}$ {\tiny (0.2--2.0~keV)} &1.4\,$\times 10^{26}$ & 2.2\,$\times 10^{27}$ & 4.8\,$\times 10^{27}$ &erg/s  \\
L$_{\rm X}$ {\tiny (2.0--5.0~keV)} & $< 10^{24}$ & 0.6\,$\times 10^{27}$ & 1.8\,$\times 10^{27}$ &erg/s  \\\hline
\end{tabular}
\end{center}
\end{table}

The spectrum of the quasi-quiescent phase is described by a moderately active corona with dominant plasma
components residing at temperatures in the range of 1.5\,--\,5.0~MK. 
The corresponding X-ray luminosity in the 0.2\,--\,2.0~keV band is $1.4 \times 10^{26}$~erg/s, 
for the ROSAT 0.1\,--\,2.4~keV band we obtain a roughly 20\,\% higher flux.
Adopting our modeling results, we derive an quasi-quiescent activity level of
$\log L_{\rm X}$/$L_{\rm bol}= -3.8$, i.e. a level not too far from the 
saturation level around $\log L_{\rm X}$/$L_{\rm bol} = -3$ for late-type stars, but much higher than those of
weakly active stars like the Sun which have $\log L_{\rm X}$/$L_{\rm bol} \approx -7$.

With its X-ray luminosity and activity level SCR~1845 is intermediate to LHS~2065 and
1RXS J115928.5-524717, two M\,9 dwarfs that are the latest X-ray detected dwarf stars \citep{rob08a,rob09}.
Given the age estimate of a few Gyr for SCR~1845, high activity levels appear as long-lasting phenomena for very low mass stars.
In comparison to activity levels of solar-like stars of the same age \citep{gue97}, SCR~1845 is hundred times more active.
Since the local stellar population has an age-spread from being rather young (few tens Myr) to intermediately old (few Gyr) and 
assuming that SCR~1845 is not atypical and its the age estimate is correct,
high activity levels are expected for a relatively large number of very low mass stars at the end of the stellar main sequence.

\subsection{Flare analysis}

To investigate the flare and its generating structure in more detail, 
we study the evolution of its X-ray emission under the assumption of a single loop as origin of the flare.
One can in principle determine the loop length by deriving the
trajectory of the flare decay in the temperature/density diagram and
adopting the formalism of \cite{rea04}.
However, the decay path can only be poorly constrained with our data, but we obtain a correction factor of the order of unity.
Alternatively we derive the loop half length $L_{9}$, formally its
upper limit since we are neglecting reheating, from the decay time $\tau$ and flare peak temperature $T_{7}$ by using
$L_{9}=\frac{\tau \sqrt{T_{7}}}{120}$.
We define the flare peak as time interval where the PN count rate exceeds 1.2~cts/s, providing
300~counts for a spectral analysis (see Tab.\,\ref{fit}).
Using our best fit values, we derive a loop half length of about $L \approx 2 \times 10^{9}$~cm, thus
the loop is clearly smaller than typical stellar dimensions.
Adopting a stellar radius of R\,$\approx$\,0.1~R$_{\odot}$, a typical value for very low-mass stars,
the loop half length is at most 0.3~R$_{*}$, i.e. it is clearly compact.

During the flare maximum SCR~1845 has an X-ray luminosity of $L_{\rm X} \approx 8 \times 10^{27}$~erg/s 
in the 0.2\,--\,5.0~keV band,
nearly sixty times above the quasi-quiescent level and plasma temperatures of up to
25\,--\,30~MK are observed during this phase. The total energy release of the flare is $1.7\times 10^{30}$~ergs for the
0.2\,--\,5.0~keV band, whereas about 30\% are emitted at energies above 2.0~keV.

We can further estimate the energy released in the U-band by using a standard count rate to flux conversion factor
($1.94 \times 10^{-16}$erg\,cm$^{-2}$\,s$^{-1}$\,\AA$^{-1}$).
When attributing the excess flux of the OM exposure no.~4 exclusively to the flare
we obtain an energy release in the U-band of about $2\times 10^{29}$~erg, i.e. a factor ten below the energy
emitted in X-rays. We note that this result is robust and we obtain similar values when using e.g. 
the Vega reference method for flux conversion.

\section{Flaring X-ray emission from ultracool dwarfs}
\label{flare}

In this section we discuss flare properties of very low mass stars in the solar neighborhood.
We focus on likely stellar objects with with spectral type M\,8 and later to have a clear separation to early/mid M~dwarfs
and require at least decent X-ray counting statistics.
In the first chapter we concentrate on active stars, i.e. objects that are similar to SCR~1845 and in the second chapter we make a comparison to the
moderately active dwarf VB~10. We note that with the exception of the L~dwarf binary Kelu-1 \citep[four photons, see][]{aud07},
no object beyond spectral type M9 has been detected in X-rays so far.

\subsection{Flares on active very low mass stars}

\begin{figure}[ht]
\includegraphics[width=92mm]{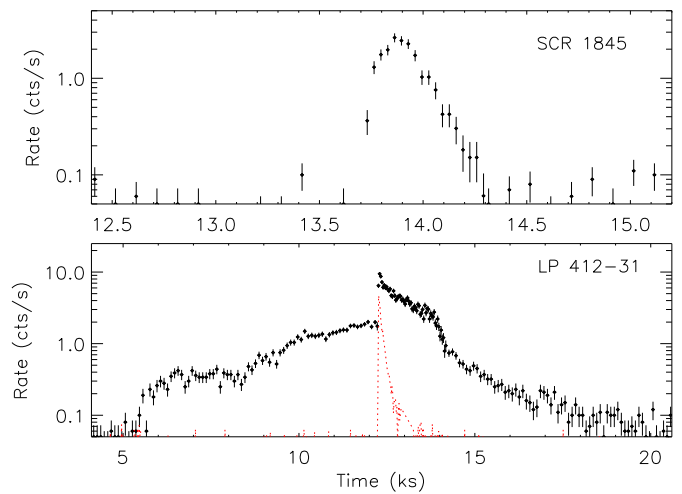}
\caption{\label{lc2}Light curves of flares on SCR~1845 (black: X-rays) and LP~412-31
(black: X-rays, red dotted: scaled OM V-band).}
\end{figure}

Only two larger flares with flux increases of at least one order of magnitude 
have been covered by modern X-ray missions an active, i.e. $\log L_{\rm X}$/$L_{\rm bol}\approx -3~...~4$, ultracool dwarfs.
In comparison to the flare on SCR~1845, the flare on the young M\,8 dwarf LP~412-31 also observed by {\it XMM-Newton} \citep{ste06},
is more energetic, has a much longer duration and exhibits a more structured X-ray
light curve as shown in Fig.\,\ref{lc2}.

The fast decay of the SCR~1845 flare
can be described by a single exponential and its origin is very likely in a single flaring loop that
emerges out of the quasi-quiescent state.
In contrast, the LP~412-31 flare shows preceding 'activity terraces' that give rise to the main flare event.
Only the very initial decay phase of the LP~412-31 flare ($\tau \approx 150\,s$) resembles the SCR~1845 decay,
later the LP~412-31 event shows a significant flattening and substructure.
Thus the flare does not decay in about 1~ks as extrapolated from the initial fast decay,
it takes roughly 7~ks before the quasi-quiescent level is reached.
This significantly longer timescale is more comparable to the one of the 
build-up of the enhanced pre-flare activity and the giant flare appears rather as the climax of an activity outbreak
that likely involves a conglomerate of magnetic structures.
Note however, that the optical emission originates nearly exclusively from the main event.
Altogether this indicates that the LP~412-31 flare loop is connected to a more complex active region, while the SCR~1845
flare originated in a more simple, isolated structure.

All active ultracool dwarf stars show
quasi-quiescent variability pointing to persistent minor activity, a 
phenomenon that is also observed on active M~dwarfs of earlier spectral type \citep[see e.g.][]{gue04a,rob05}.
However, the flare amplitude distribution might be different as already noted by \cite{ste06} for LP~412-31,
a peculiarity that applies similarly to SCR~1845.
Very low mass stars seem to release magnetic energy more preferably in singular, larger events 
rather than in the canonical flare distribution observed on the Sun and other cool stars, 
where the energy-release follows a power law with an exponent of $\approx -2$ \citep[see e.g.][]{aud00,gue04}. 
To have a complete coverage of available observations of suitable targets, 
we further investigated another archival 40~ks observation of LP-412-31 with {\it Chandra} and find that is
shows exclusively quasi-quiescent emission without flaring. These data
confirm it to be a very active star with a strong hot (10~MK)
plasma component in quasi-quiescence, albeit with $L_{\rm X}= 6 \times 10^{26}$~erg/s in the 0.2\,--\,2.0~keV band it has a factor two lower 
X-ray luminosity compared to
the {\it XMM-Newton} observation performed nearly two years before.
Together with the above mentioned observations of the M9 dwarfs LHS~2065 and RXS~1159,
in total nearly 2~days (155~ks) of X-ray observations of active ultracool dwarf stars have been performed.
This resulted in the observation of two large flares ($L_{\rm X\,\it flare} \gtrsim 50 \times L_{\rm X\,\it qq}$) 
and frequent minor variability of a factor up to a few, but no intermediate events.

An explanation may lie in the combination of the convective stellar structure and the
resistivity of the outer atmospheric layers; additionally an efficient dynamo mechanisms
must be present. Phenomenologically one might speculate that
it requires a significant amount of magnetic energy to break through from the stellar interior producing the large flares, 
whereas the frequent, smaller variability might be related to magnetic activity that
is generated closer to the stellar surface and released regularly. An underlying dynamo could be of the $\alpha^{2}$ or 
the turbulent type \citep[see e.g.][]{cha06, dur93}, but likewise both mechanisms may be operational at the same time.
Anyway, despite different underlying stars, energetics and light curve evolution,
both flares are isolated, vigorous events in a vastness of quasi-quiescent X-ray emission.

\subsection{The case of VB 10}

We compare our findings to the prominent, moderately active M\,8 dwarf VB~10.
It is located at a distance of 6.1~pc and has an X-ray activity level of $\log L_{\rm X}$/$L_{\rm bol}\approx -5$,
using $L_{\rm bol}= 1.7 \times 10^{30}$~erg/s from \cite{fle03}.
Thus it is one to two magnitudes less active than the above discussed objects.
It has been observed in X-rays by several missions, e.g. two {\it Chandra} ACIS-S observations are published in the literature \citep{fle03, ber08b};
theses works contain also a more detailed description of this object.
VB~10 is the only moderately active ultracool dwarf star where more that a few X-ray photons have been detected so far.

The first {\it Chandra} observation (Obs.ID 616) performed in July 2000 with a duration of 13~ks leaded to the
detection of 26 counts on the back-illuminated S3~chip; no obvious strong variability is present and \cite{fle03} modelled the 'quiescent corona'
with a 3~MK plasma, derived an X-ray brightness of $L_{\rm X}= 2.4 \times 10^{25}$~erg/s
and an activity level of $\log L_{\rm X}$/$L_{\rm bol}= -4.9$. The second {\it Chandra} observation (Obs.ID 7428) with a duration of 29~ks contains 
58 counts, i.e. on average a very similar photon flux \citep{ber08b}. 
However, in contrast to what is stated in their work, during this observation VB~10 is not located 
on the back-illuminated S3~chip but on the front-illuminated S2 chip.
The front-illuminated chips have significantly less effective area at lower energies and thus the count rates and light curves are not directly comparable.
Nevertheless, more emission from hotter plasma is present in July 2007, as deduced from spectral modelling 
and attributed by the authors to two flares on an underlying quiescent phase.
They derived a factor two higher X-ray luminosity of $L_{\rm X}= 5.4 \times 10^{25}$~erg/s
from a two temperature model with a 3~MK and 15~MK component and an emission measure ratio of roughly two to one.
In this observation VB~10 is, compared to the average flux, by a factor of two to three moderately brighter
in the beginning and correspondingly fainter at later times.
While clearly X-ray variability at a level of a few up to an order of magnitude
is present, a detailed investigation, especially of individual flares, suffers from the low count rate.

\begin{figure}[ht]
\hspace*{2mm}\includegraphics[width=90mm]{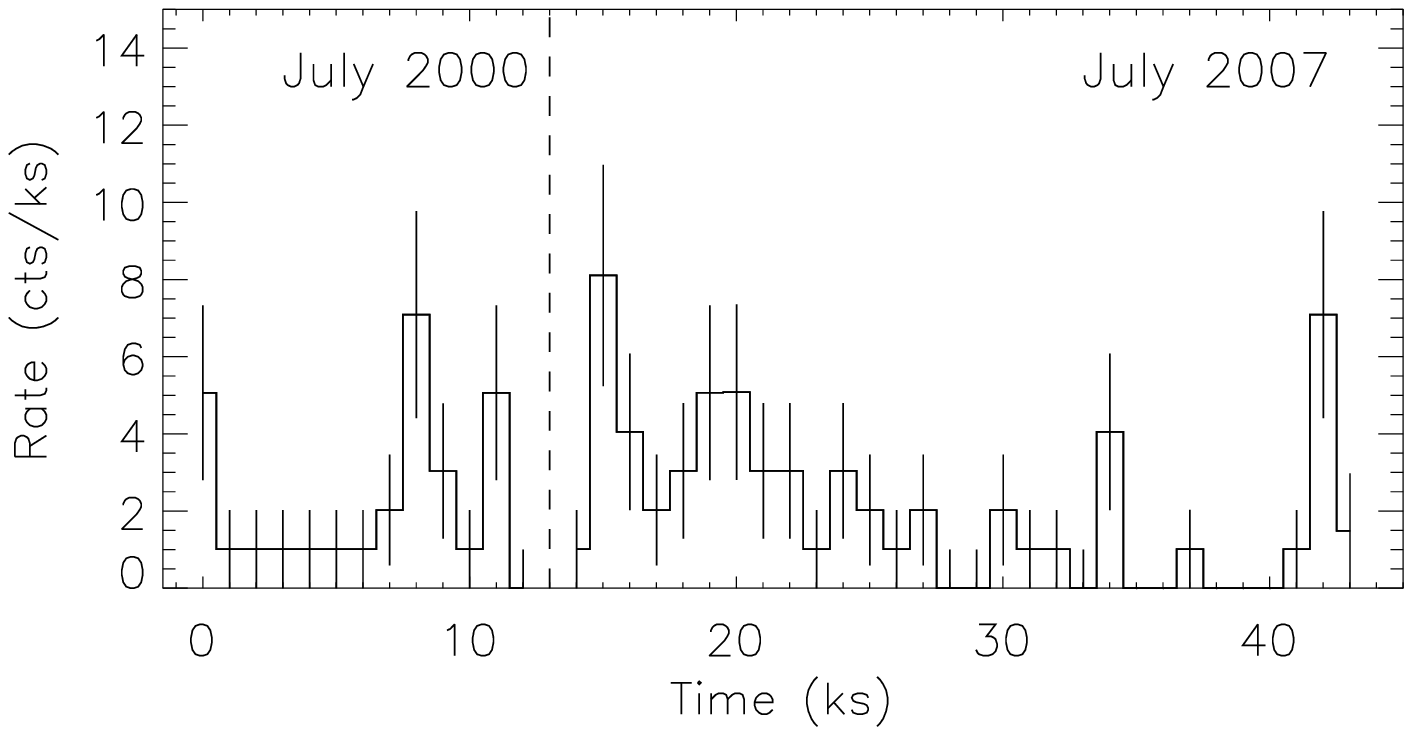}
\vspace*{1.4mm}

\hspace*{2mm}\includegraphics[width=90mm]{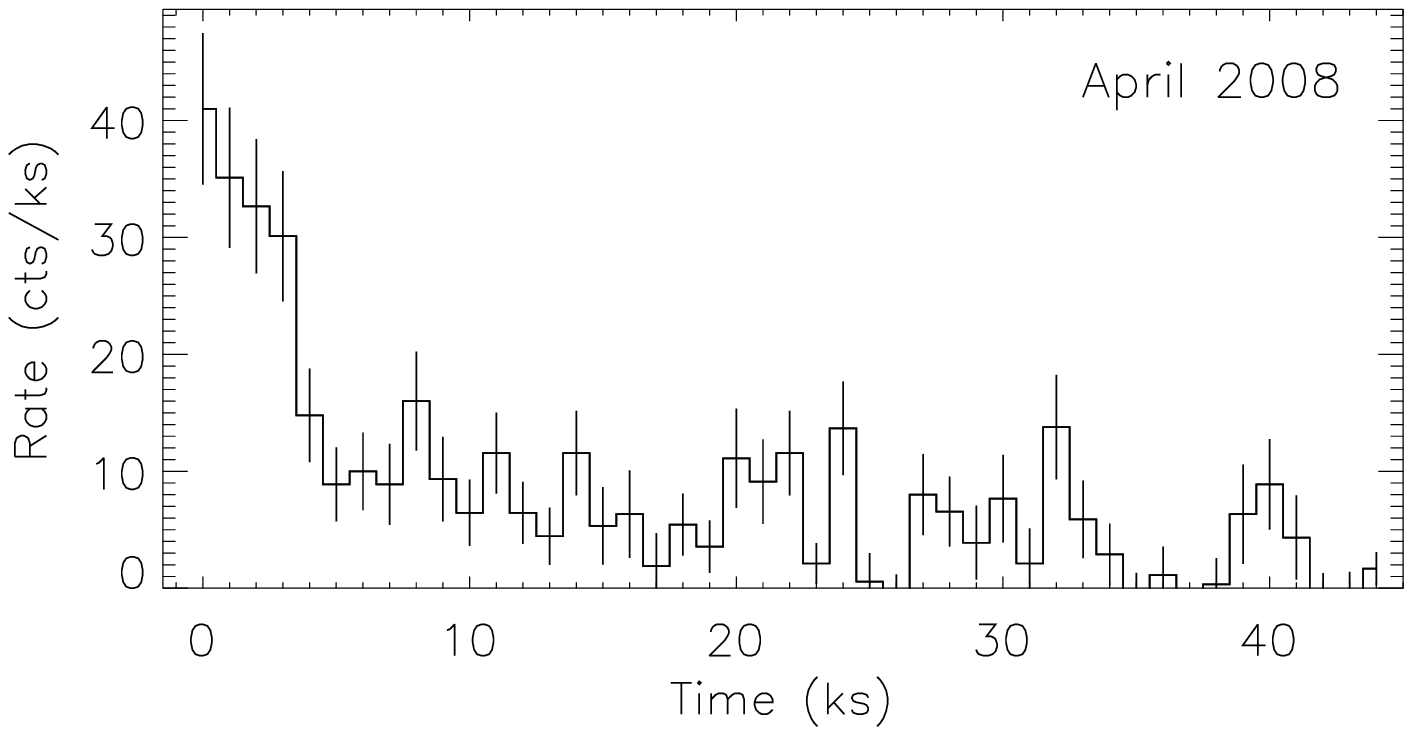}
\vspace*{1.5mm}
\caption{\label{vb10lc}X-ray light curves of VB~10 with 1~ks binning 
{\it Upper panel}: Chandra, ACIS-S (2000: BI-chip, 2007 FI-chip), 0.15\,--\,2.0~keV. {\it Lower panel}:
XMM-Newton, MOS\,1+2, 0.2\,--\,2.0~keV.}
\end{figure}

In Fig.\,\ref{vb10lc} we
show X-ray light curves of VB~10 uniformly binned to 1~ks. The upper panel shows the light curves
obtained from a re-analysis of the {\it Chandra} 
datasets with source photons extracted from a circular region with 2\arcsec\,radius in the energy band 0.15\,--\,2.0~keV.
Overall, the X-ray light curves of the July 2000 phase and the July 2007 phase look quite similar, but remind that different CCD chips were used.
To derive a more comprehensive picture of VB~10, we additionally investigate a dataset from {\it XMM-Newton} (Obs.ID 0504010101), 
that we retrieved from the archive.

The {\it XMM-Newton} observation is with 44~ks (MOS detectors) the longest of the three and the usage of the thin filter
provides high sensitivity at low energies, however it is affected by time intervals of high background.
Nevertheless we are able to extract basic X-ray properties for VB~10, utilizing similar selection criteria as described above for SCR~1845.
The XMM observation contains a clear example of an individual flare on VB~10 (Fig.\,\ref{vb10lc}, lower panel), but
unfortunately the peak is not covered and only the decay is partially observed; we note that
the more sensitive PN detector started observing even later. The shown MOS light curve is background subtracted and was extracted from
a 15\arcsec\,circular region in the 0.2\,--\,2.0~keV band.

Beside the flare at the beginning of the observation additional variability is present over the whole 
XMM exposure, however the exact level is
again poorly constrained. The overall X-ray spectrum is well described by a two temperature model with 
plasma components at 2.5~MK and 7.5~MK and an emission measure ratio of about two to one, quite similar to the {\it Chandra}
observation performed in July 2007. We derive an average X-ray luminosity of $L_{\rm X} = 1.2 \times 10^{26}$~erg/s and for the
flare phase (0\,--\,8~ks) $L_{\rm X} = 5.4 \times 10^{26}$~erg/s. For the remaining 
quasi-quiescent phase we obtain $L_{\rm X} = 7.4 \times 10^{25}$~erg/s; 
here already a 3~MK one temperature model describes the data quite well. When
considering only the last 20~ks of the observation, i.e. the phase sufficiently after the flare event, 
we obtain an X-ray brightness of $L_{\rm X} = 3.2 \times 10^{25}$~erg/s, very
similar to VB~10 in the July 2000 observation. Altogether, the variability observed within half a 
day by {\it XMM-Newton} mainly covers the states of VB~10 observed in two {\it Chandra} exposures separated by seven years, with an additional flare
being caught in its decay phase. This points to long term stable but always moderately variable X-ray emission, a state that we
prefer to name - given the sensitivity of present day X-ray instrumentation - quasi-quiescence.

We note the detection of a larger X-ray flare on VB~10 with ROSAT HRI (10 photons), 
reaching an average X-ray brightness in the order of $L_{\rm X} = 10^{27}$~erg/s and an
at least three times higher peak flux \citep{fle00}, implying a flux increase by a factor of 100.
Thus also moderately active ultracool dwarf stars produce intense X-ray flares, however a detailed study of time variability
is complicated by their intrinsic X-ray faintness.

\section{Summary and conclusions}
\label{sum}

   \begin{enumerate}
\item We detect quasi-quiescent X-ray emission as well as a large flare from the nearby, ultracool M8.5/T dwarf binary SCR~1845. 
This is the first report of magnetic activity from this object.

\item In the quasi-quiescence state SCR~1845 is detected at soft X-ray energies below 2.0~keV
with an X-ray luminosity of $L_{\rm X} = 1.4 \times 10^{26}$~erg/s. This leads
to an activity level of log~$L_{\rm X}$/$L_{\rm bol}\approx -3.8$, pointing to a rather active star
despite its age of several Gyr.
SCR~1845 is one of the coolest main sequence star with securely detected quasi-quiescent X-ray emission.

\item The flare has short duration of about 10~min and outshines the parent star by a factor of 50 at X-ray energies 
with a peak X-ray luminosity of log~$L_{\rm X} =27.9$~erg/s. Flare temperatures reach about
30~MK and in total roughly $2\times 10^{30}$~erg are emitted at X-ray energies during the event. 
The flare originates from a rather compact structure and
is accompanied by a significant NUV/optical brightening in the corresponding OM image.

\item The X-ray data from SCR~1845 and other recent observations support the hypothesis, that the flare energy distribution in active very low mass stars
differs from the one observed in more massive late-type stars.

   \end{enumerate}

\begin{acknowledgements}
This work is based on observations obtained with XMM-Newton, an ESA science
mission with instruments and contributions directly funded by ESA Member
States and the USA (NASA) and makes use of data obtained from the Chandra data archive.
J.R. and K.P. acknowledge support from DLR under 50QR0803 and 50OR0703 respectively.

\end{acknowledgements}

\bibliographystyle{aa}
\bibliography{13603}

\end{document}